# Quantifying the Temperature of Heated Microdevices Using Scanning Thermal Probes

*Amin Reihani [†, 1], Shen Yan[†, 1], Yuxuan Luan[†, 1], Rohith Mittapally[1],*

*Edgar Meyhofer[*, 1], Pramod Reddy[*, 1, 2]*

[1]Department of Mechanical Engineering, University of Michigan, Ann Arbor, MI 48109

[2]Department of Materials Science and Engineering, University of Michigan, Ann Arbor, MI 48109

[†]These authors contributed equally to this paper.

*Correspondence to be addressed to, email: meyhofer@umich.edu, pramodr@umich.edu*

**Abstract**

Quantifying the temperature of microdevices is critical for probing nanoscale energy transport. Such quantification is often accomplished by integrating resistance thermometers into microdevices. However, such thermometers frequently become structurally unstable and fail due to thermal stresses at elevated temperatures. Here, we show that custom-fabricated scanning thermal probes (STPs) with a sharp tip and an integrated heater/thermometer can accurately measure the temperature of microdevices held at elevated temperatures. This measurement is accomplished by introducing a modulated heat input to the STP after contacting the microdevice with the STP's tip, and characterizing the DC and AC components of the STP's temperature. From these measured temperature components, the tip-to-sample thermal resistance and the microdevice surface temperature are deduced via a simple lumped-capacitance model. The advances presented here can greatly facilitate temperature measurements of a variety of heated microdevices.

Quantifying the temperature of heated microscale devices is critical for successfully employing them in probing nanoscale transport via conduction and radiation. Such quantification becomes challenging for small devices at high temperatures due to the difficulties in integrating reliable temperature sensors. Past work for quantifying high temperatures at the micro/nanoscale has relied on employing approaches such as phase change of nanoparticles[1], Raman spectroscopy[2], complicated approaches involving integration of fine thermocouples[3] that lead to destruction of the microdevice rendering them useless for the intended studies and other optical approaches[4]. Typical examples requiring the characterization of the temperature of microdevices are related to the exploration of near-field thermal transport[5,6] and near-field based thermophotovoltaic energy conversion[3,7] where heated microdevices are routinely employed. In this work we show that scanning thermal probes (STPs), which have been recently employed for probing a number of interesting thermal phenomena[8-18], can be used to quantitatively measure the temperature of heated microdevices (referred as also a device under test (DUT)) in a non-destructive fashion.

Our approach to quantify the temperature of heated DUTs relies on custom-fabricated STPs (see Fig. 1(a)) with "I-shaped" beams that enable both a relatively large mechanical stiffness and a low thermal conductance (see supporting information (SI)). These probes also feature an integrated Pt resistance thermometer, which allows both temperature sensing and heating of the STP (see SI). In our method, we measure the thermal response of the STP to heat inputs in two *scenarios*: i) when the STP is in contact with the DUT (see Fig. 1(c)) and ii) when it is far away from the DUT. As explained in more detail below, the measured thermal responses in these two conditions enable us to quantify the temperature of the heated DUT.

The thermal behavior of the STP under unmodulated (DC) and modulated (AC) heat inputs can be described by a dynamic lumped-capacitance thermal model[19], as represented by the thermal networks illustrated in Figs. 1 (d) and (e), respectively. In these figures, $T_S$ represents the surface temperature of the DUT that needs to be quantified, and $T_R$ and $T_{tip}$ represent the thermal reservoir and probe tip temperatures. The thermal capacitance of the probe, $C_{probe}$, is lumped in the node $T_{tip}$. In addition, $R_C$ and $R_{TS}$ represent the thermal resistances of the probe cantilever and tip-sample contact. The parameters $C_{probe}$, and $R_C$ depend on temperature-dependent material properties and are obtained from a set of calibrations. Key to quantitative measurement of $T_S$ is a direct measurement of $R_{TS}$, which depends on the detailed geometry and the local thermal and mechanical properties of the tip at the point of contact and can vary spatially and temporally as the tip-sample geometry and mechanics change. Below, we explain how $R_{TS}$ is characterized via a modulated heat input to the probe.

We first note that the STP can be heated by supplying an AC current at frequency $f_1$ to the probe heater, which results in an AC heat input at frequency $2f_1$ as well as a DC heat input via Joule heating. Further, the laser employed to detect the deflections of the STP introduces additional DC heating. Hence, the total heat input to the STP has a DC component ($\dot{q}_{DC}$) and AC component ($\dot{q}_{AC}$), resulting in DC and AC temperature changes of the probe (see SI for details).

To quantify the values of $R_C$ and $C_{probe}$ we characterized the thermal frequency response of our STP. These experiments were performed at various DC temperatures ($T_{DC}$) achieved by varying the incident laser power on a free standing STP (i.e. the STP is not in contact with the DUT). At each $T_{DC}$ a modulated heat input $\dot{q}_{AC}$ = 6.1 µW at various frequencies was applied to the STP heater. The resulting temperature oscillations of the STP ($T_{AC}$) are shown in Fig. 2(a) as distinct curves at each DC temperature. Using the low frequency (0.5 Hz) data, the thermal conductance of the probe ($G_C$ = 1/$R_C$) was estimated for various values of $T_{DC}$ as shown in Fig. 2(b). It can be seen that the probe conductance increases as a function of temperature, which can be attributed partly to an increase in thermal conductivity[20] of $SiN_x$, from which our STPs are made, and partly to the radiative conductance of the membranes (which feature large absorption cross-sections[21-23]) of the STP that is expected to increase with temperature. Finally, Fig. 2(c) shows the thermal cutoff frequency (-3 dB point) that is weakly dependent on temperature, from which one can estimate the thermal capacitance of the probe ($C_{probe} = G_C/2\pi f_C$).

In order to quantify $R_{TS}$, we performed an experiment where the STP cyclically contacts the sample (surface of the DUT) and moves up to a height of ~2 μm above the surface. At these gap-sizes, contributions from radiative heat transfer (especially near-field radiative heat transfer) are expected to be negligibly small[11,24] (i.e. $R_{TS} \gg R_C$)). As the STP contacts the sample and separates from it, $T_{DC}$ and $T_{AC}$ values were continuously recorded. In order to distinguish the two cases, we label the DC and AC temperatures of the STP as $T_{DC-o}$ and $T_{AC-o}$ (when the STP tip is separated from the sample), and $T_{DC-c}$ and $T_{AC-c}$ (when the probe is in contact with the sample). We note that contact of the STP with the DUT results in a change in the AC and DC components of the STP temperature. We quantify the change in AC temperature by defining the ratio $r = (T_{AC-o}/\dot{q}_{AC-o})/(T_{AC-c}/\dot{q}_{AC-c})$, where ($\dot{q}_{AC-o}$) and ($\dot{q}_{AC-c}$), are the AC heat inputs before and after contact, respectively. These two heat inputs are slightly different due to a change in the electrical resistance of the heater upon contact. One can calculate $R_{tot}$ (defined as $R_{tot}^{-1} = R_C^{-1} + R_{TS}^{-1}$) using the measured $r$ and the relation below where $\omega = 4\pi f_1$ is the angular frequency of AC heat input to the STP (see SI for full derivation).

$$R_{tot} = \sqrt{\frac{R_C(T_{DC-o})^2}{r^2\left(1+R_C(T_{DC-o})^2 C_{probe}(T_{DC-o})^2\omega^2\right)-R_C(T_{DC-o})^2 C_{probe}(T_{DC-c})^2\omega^2}} \qquad \text{Eq. 1}$$

Subsequently, the tip sample contact resistance can be obtained from $R_{TS}^{-1} = R_{tot}^{-1} - R_{C}^{-1}$. Once $R_{TS}$ is determined, the sample temperature can be obtained from (see SI):

$$T_S = \frac{R_{TS}}{R_C}(T_{DC\text{-}c} - T_{DC\text{-}o}) + T_{DC\text{-}c} + R_{TS}(\dot{q}_{DC\text{-}o} - \dot{q}_{DC\text{-}c}) \qquad \text{Eq. 2}$$

In order to confirm the validity and accuracy of the proposed measurement technique, we employed a suspended microdevice as DUT whose temperature ($T_S$) can be increased via an integrated Pt resistance heater and also be simultaneously measured using the same Pt resistance thermometer (PRT) (Fig. 1(b), see SI for measurement details). In these studies, the STP was placed close to the center of the microdevice mesa as shown in Fig. 1(c). At each value of heating power applied to the microdevice, a sequence of approach and retract cycles were performed and $T_{DC}$, $T_{AC}$ values were recorded. Figs. 3(a) and (b) show the time series of this measurement (bandwidth of ~1 Hz) at two heating powers of the microdevice (0 and 56 mW) corresponding to surface temperatures of $T_s$ = 25.1 and 330.6 °C respectively, each showing two repeats of approach and retract cycle.

As can be seen in Fig. 3(a), $T_{DC\text{-}o}$ ≈ 45 °C, this is due to $\dot{q}_{DC}$ provided by the AFM laser as well as the DC component of heat input provided by probe heater. When the probe is placed in contact with an unheated microdevice $T_{DC\text{-}c}$ drops to ~35 °C as a result of contact to a colder surface. In the subsequent retract and contact phases the probe establishes a significantly different contact and reaches ~40 °C, which indicates the stochastic behavior of the probe-sample thermal contact resistance. This variation in $R_{TS}$ is also reflected in the independently measured $T_{AC}$, which is correspondingly lower. In Fig. 3(b), we present data from an experiment where the DUT is heated. It can be seen that $T_{DC\text{-}o}$ ≈ 50 °C when the probe is retracted, this is higher than the previous case of Fig. 3(a) due to additional far-field radiative heating of the probe by the DUT surface. During the first approach, the $T_{DC-c}$ reaches ~149 °C while during second approach it reaches ~147 °C which indicates that in this experiment the two approaches lead to similar contacts. This is also reflected in the $T_{AC}$ signal, which shows a very small difference between the two contacts.

In order to quantify the surface temperature, we use Eqs. 1, and 2, where each data point corresponding to the probe approach (shaded regions in Fig. 3) is used separately to obtain various estimates of $T_{DC-c}$, and an average of data points when probe is retracted (unshaded regions in Fig. 3) is used to estimate $T_{DC-o}$. This procedure leads to evaluation of $R_{TS}$ and $T_S$ for each data point during contact. The results from this analysis are shown in Figs. 4 (a) and (c), from which it can be seen that the $R_{TS}$ values for the first and second approach demonstrate different mean values of $0.87 \times 10^6$ K·W$^{-1}$ and $3.00 \times 10^6$ K·W$^{-1}$, indicating that different contact geometries were established during these two approaches. On the other hand, the mean values of $R_{TS}$ for the second experiment, shown in Fig. 4(b), are similar i.e. $1.08 \times 10^6$ K·W$^{-1}$ and $1.21 \times 10^6$ K·W$^{-1}$ for the first and second approach, respectively. Overall these measured values of $R_{TS}$ were found to be in good agreement with expectations (see SI) based on models from past work[25].

Histograms corresponding to the measured temperature $T_s$ from the data in Figs. 4(a), (c) are shown in Figs. 4(b), (d), respectively. From Gaussian fits to each histogram the mean values of $T_S$ in each of the experiments are estimated to be 24.5 °C and 327.0 °C with standard deviations of 0.7 K and 9.0 K, respectively. These estimates are to be compared with independent measurements of the emitter temperatures, using the thermometer embedded into the emitter, which are 25.1 °C and 330.6 °C. As can be seen, the agreement between the two measurements is good, however, at higher temperatures the uncertainty is found to increase possibly due to fluctuations in $R_{TS}$ at high temperatures.

Finally, in Fig. 5, we present experimental data from validation measurements at various surface temperatures that were achieved by varying the power dissipated in the instrumented DUT. As can be seen, the measured data from the STP are in very good agreement with DUT PRT measurements up to 330°C. Above ~370°C, the PRT resistance of the heated DUT starts to drift making the measurements unreliable. However, the measurements performed using the STP continue to correlate well with the increasing power up to 417 °C (as estimated from the STP measurements) at which point the Pt heater on the DUT failed, making it impossible to evaluate the response at even higher temperatures. It is not entirely surprising that the STP is capable of measurements at high surface temperatures as the net temperature rise of the STP is much smaller than that of the DUT due to the large tip-sample thermal resistance ($R_{TS}$). In general, a higher $R_{TS}$ value is expected to result in an increased operating temperature range of the STP at the expense of reduced temperature sensitivity. The $R_{TS}$ value can in principle be tuned for each experiment, for instance, one can significantly increase the magnitude of $R_{TS}$ by depositing a thin layer of low conductivity material on the sample or the probe (such as $Al_2O_3$).

In summary, the temperature measurement approach described here represents a direct method for measuring the temperature of microdevices that is particularly well-suited for small devices operated at high temperatures. This approach leverages the large thermal resistance of a nanoscale contact, which serves to both minimize the perturbation of the microdevice temperature and limits the temperature rise of the probe enhancing the range of temperatures that can be measured. The agreement between the temperatures measured with the STP and those with the Pt thermometer integrated into the microdevice supports the robustness of the approach reported in this paper. We expect that the developed method can greatly help in characterizing microscale devices employed in a range of nanoscale energy transport experiments.

**Acknowledgements**

We acknowledge support from ONR under award No. N00014-16-1-2672 (experimental work), DOE-BES through a grant from the Scanning Probe Microscopy Division under award No. DE-SC0004871 (fabrication of STPs), support from and support from the Army Research Office under award No. W911NF-19-1-0279 (fabrication of emitters).

**Data Availability**

The data that support the findings of this study are available from the corresponding authors upon reasonable request.

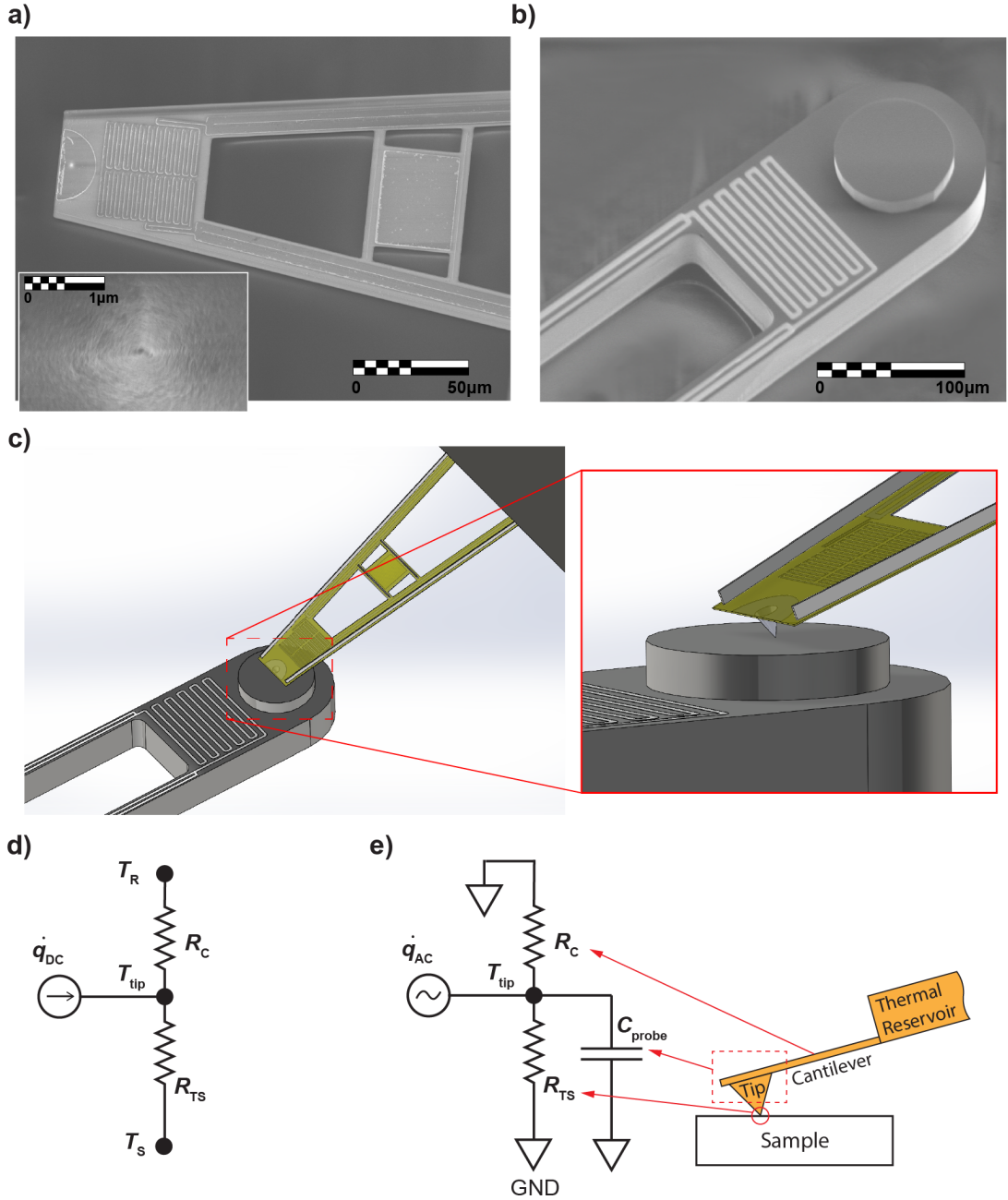

**Fig 1. (a)** Scanning electron micrograph (SEM) of the STP probe with an integrated Pt heater thermometer and I-shaped beams. The inset shows the STP tip. **(b)** SEM of the suspended microdevice (DUT) with embedded Pt heater. **(c)** Schematic illustrating the geometry of the STP in contact with heated DUT. Note, to obtain a clear view, the tip height is exaggerated. **(d)** DC- and **(e)** AC-equivalent thermal resistance network corresponding to the geometry shown in (c). $T_S$, $R_C$ and $R_{TS}$ represent the DUT temperature, thermal resistance corresponding to the beams of the STP and tip-sample thermal contact resistance, respectively. While $C_{probe}$, $\dot{q}_{DC}$ and $\dot{q}_{AC}$ represent the lumped thermal capacitance of the STP, and the DC and AC components of the heat input to the STP, respectively.

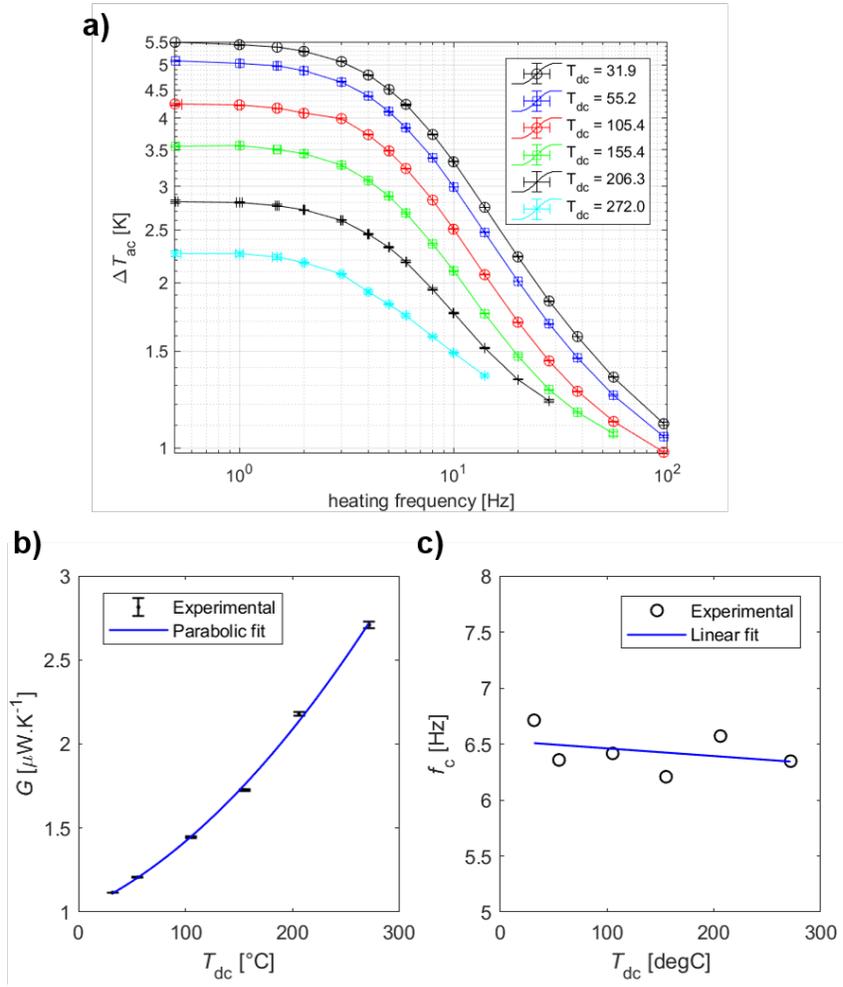

**Fig 2. (a)** Frequency response of a STP probe that is located away (10 μm from the sample surface) measured at various DC temperatures established by varying the laser power incident on the STP. All data point obtained at constant heating amplitude of $\dot{q}_{AC}$ = 6.1 μW. **(b)** Thermal conductance of the probe at the low-frequency limit (0.5 Hz) as a function of probe temperature. **(c)** Thermal cut-off frequency of the STP probe based on the -3dB point as a function of probe temperature.

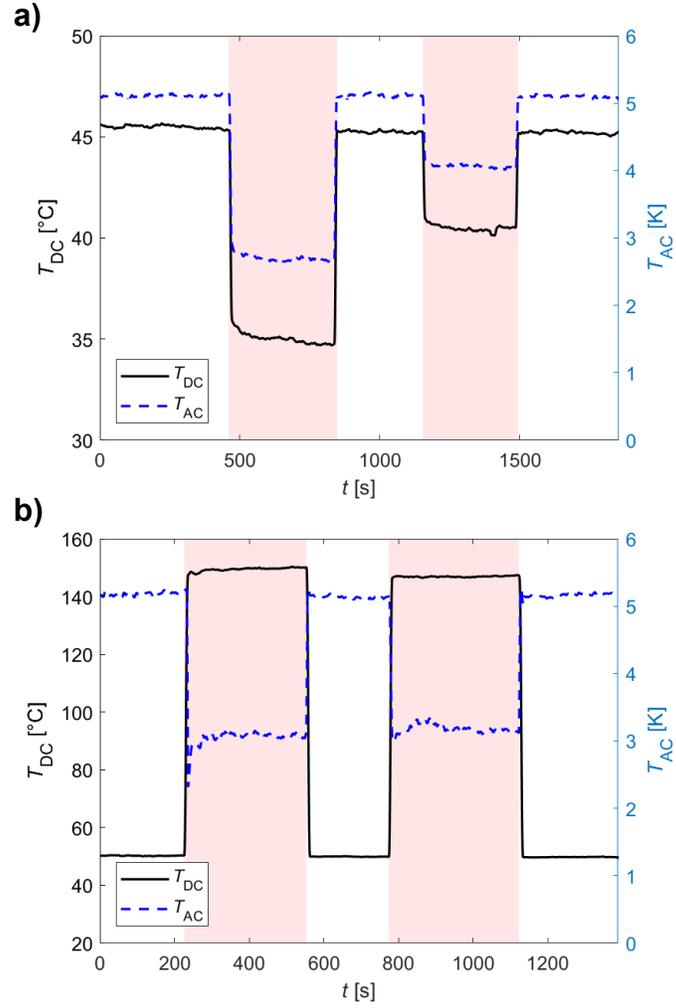

**Fig 3.** Time series of DC and AC temperatures at the tip of STP tip ($T_{tip}$) during two successive contact and retraction cycles on the silicon mesa of the heated microdevice. The surface temperatures are chosen to be $T_S$ = 25.1 °C for the experiment shown in **(a)** and $T_S$ = 330.6 °C for the experiment show in **(b).**

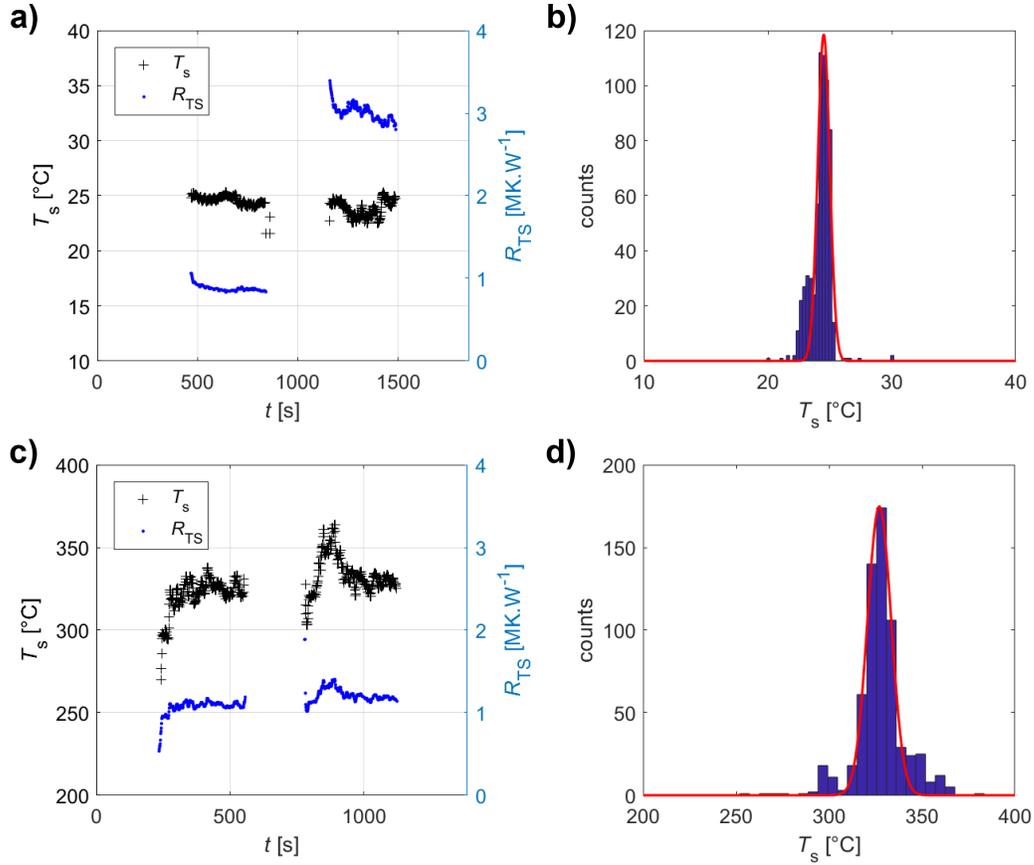

**Fig 4.** Values of surface temperature ($T_S$) and tip-sample thermal contact resistance ($R_{TS}$) estimated from the data shown in Fig. 3. For the data shown in **(a)**, the surface temperature of the microdevice ($T_S$) was set to be 25.1 °C, whereas for the data shown in **(c)**, it was chosen to be 330.6 °C. **(b)** Histogram constructed from the $T_S$ data shown in **(a)**. **(d)** Same as **(b)**, but for data shown in **(c)**. Gaussian fits to the data in **(c)** and **(d)** shown in red.

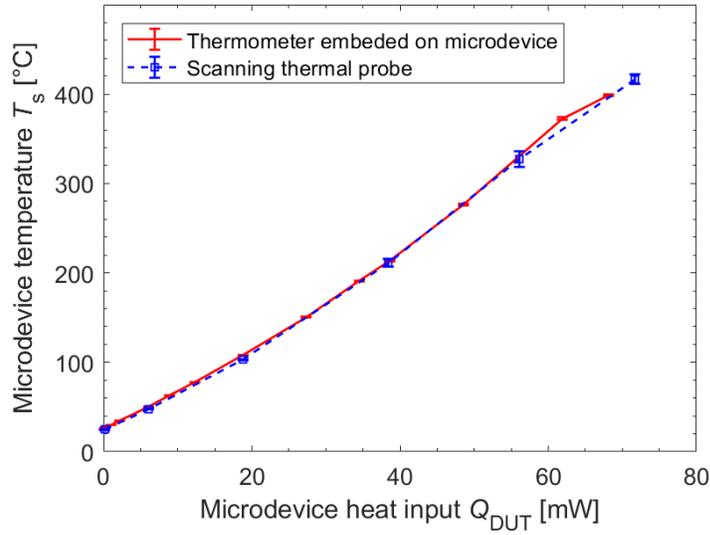

**Fig 5.** Measured temperature ($T_S$) of heated microdevice (DUT) using the STP (blue) and a Pt resistance thermometer (red) integrated into the heated microdevice as a function of power dissipated in the microdevice ($Q_{DUT}$). Note, $T_S$ is not a linear function of $Q_{DUT}$ because the thermal conductance of the heated microdevice decreases from 238 µW·K$^{-1}$ to 178 µW·K$^{-1}$ as the temperature is increased from 25 to 372 °C. The error bars correspond to ±1 standard deviation as estimated from the Gaussian fits to the histograms shown in Fig. 4.

# Supporting Information

# Quantifying the Temperature of Microdevices Using Scanning Thermal Probes

*Amin Reihani [†, 1], Shen Yan[†, 1], Yuxuan Luan[†, 1], Rohith Mittapally[1],*

*Edgar Meyhofer[*,1], Pramod Reddy[*,1, 2]*

[1]*Department of Mechanical Engineering, University of Michigan, Ann Arbor, MI 48109*

[2]*Department of Materials Science and Engineering, University of Michigan, Ann Arbor, MI 48109*

[†]These authors contributed equally to this paper.

*Correspondence to be addressed to, email: meyhofer@umich.edu, pramodr@umich.edu*

## S1. Scanning Thermal Probe Fabrication

The steps involved in the fabrication of the STPs employed in this work are shown in Fig. S1. In brief, a T-beam was first defined on a bare Si wafer through deep reactive ion etching (DRIE), followed by wet oxidation and low pressure chemical vapor deposition (LPCVD) of 600 nm silicon nitride. The T-beams improve the stiffness of the probe cantilever which enable stablishing stable thermal contacts to the sample during measurements[1]. After that, the back side of the wafer was patterned by reactive ion etching (RIE). Subsequently, a 3 μm thick low temperature oxide (LTO) was deposited with Cr caps defined on the front side to form conical tips. Then, LTO was etched by buffered HF (BHF), leaving a tip of 3 μm height under the region where the Cr cap was located. Afterwards, the tip was coated by a 150 nm Cr layer. Next, a 50 nm thick serpentine Pt line was deposited for creating a heater-thermometer for resistive heating and thermometry. Later, a Au reflector and electrode were patterned. Then the front side of the wafer was covered by as 50 nm thick plasma enhanced chemical vapor deposition (PECVD) nitride layer to hinder the etching rate at the front side. Finally, the probe was released by KOH etching of Si.

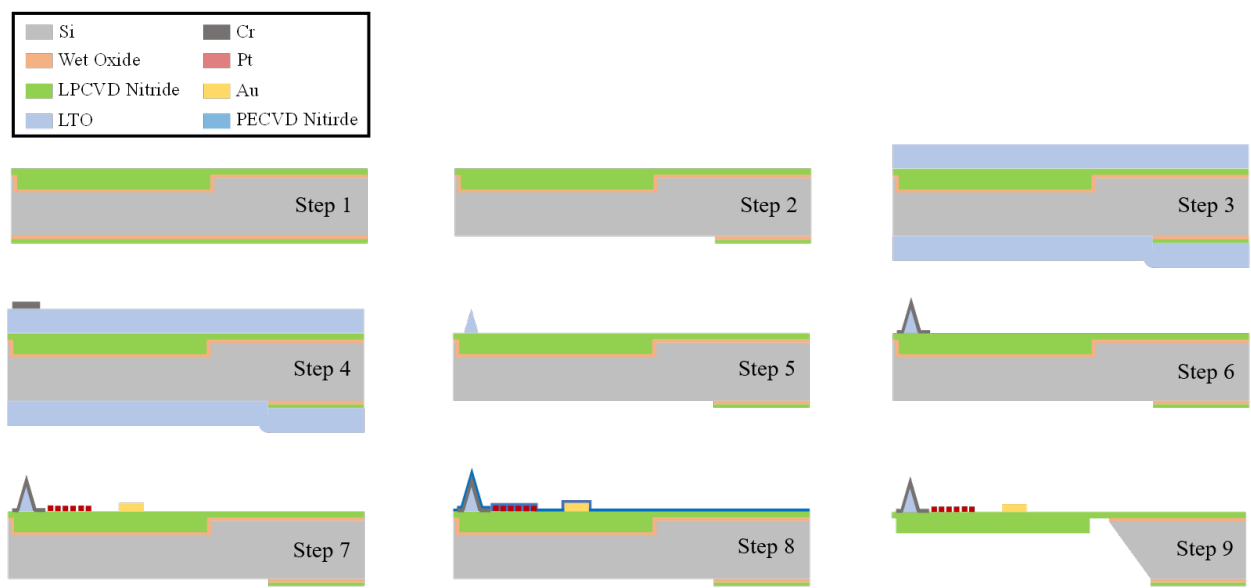

*Fig. S1* *Process steps for fabrication STPs with integrated platinum resistive heater/thermometer.*

## S2. Finite Element Modeling of Scanning Thermal Probe

We estimated the mechanical stiffness of the probe by performing finite element simulations in COMSOL$^{TM}$. Both the geometry of the probe and various layers of Au, Pt and Cr were included in the model. The probe consisted of 280 μm long, 600 nm thick SiN$_x$ cantilever with thin films of Pt (50 nm), Au (100 nm) and Cr (150 nm) on it. A force ($F$) of 1 μN was applied to the tip end of probe body while the other end of probe was fixed. Tip deflection ($\delta$) was calculated under the applied force and the stiffness of the probe ($k = F/\delta$) was estimated to be 1.79 N/m as shown in Fig. S2 (a).

The thermal conductance of the probe was also studied using the same model. For the SiN$_x$ thin films, a thermal conductivity of 5.0 W/m-K was assumed[2]. A known heat flux was applied to the tip end of the probe while the temperature of the other end was fixed at room temperature (293.15 K), generating a temperature gradient along the probe body. Tip temperature rise was monitored under various input heat fluxes and the thermal conductance of the probe was extracted by linearly fitting the data to be 0.69 µW/K close to room temperature as shown in Fig. S2 (b). Note that the discrepancy observed between the model and the experimentally obtained value (1.11 µW/K at 31.9 °C from Fig. 2 (c) of the manuscript) can be attributed to the variation in SiN$_x$ thin film thermal conductivity from the assumed value (0.69 µW/K), differences in the location of heat input (tip end in the model, distributed along the Pt serpentine in experiments), and contributions from thermal radiation.

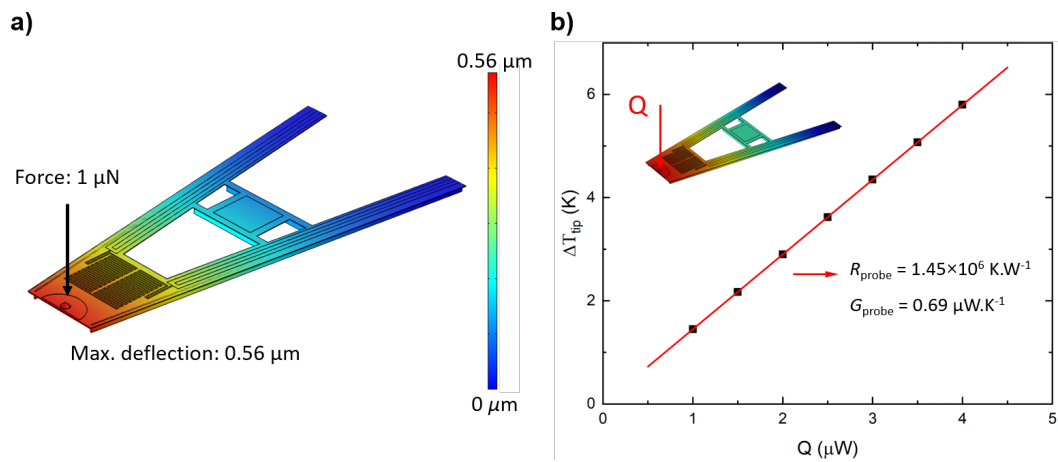

*Fig. S2 (a) Finite element simulation of STP deflection under an applied load. (b) Temperature rise of the probe as a function of heat input to the tip, the slope of the linear fit shows the probe thermal resistance ($R_{probe} = 1/G_{probe}$).*

## S3. Scheme Employed for Heating and Measuring Temperatures of STPs

A platinum resistive thermometer (PRT) with 8 kΩ nominal resistance was integrated into the end of the STP. Temperature sensing was performed by reading the resistance of the PRT using a full Wheatstone bridge[3] via a multi-frequency lock-in approach as described below. The bridge was driven by a current source with a waveform which was a superposition of two sine waves as $I(t) = A_1 \sin(2\pi f_1 t) + A_2 \sin(2\pi f_2 t)$. The first sine wave had a low frequency ($f_1$ = 1 Hz) and high amplitude ($A_1$ = 56 µA) which was used for modulated heating of the probe. Since the joule heating frequency $2f_1$ = 2 Hz was smaller than the thermal cutoff frequency (~6.5 Hz) efficient heating of the probe was achieved. The modulated temperature amplitude of the probe, $T_{AC}$, was measured by lock-in technique using the $3f_1$ component of the Wheatstone bridge output[4]. The second sine wave had a high frequency ($f_2$=200 Hz) and lower amplitude ($A_2$ = 14 µA), which was used for readout of DC temperature of the probe by lock-in technique using the $f_2$ component of the Wheatstone bridge output.

## S4. Heating and Temperature Sensing of the Heated Microdevice Sample

A microdevice used as a heated sample for verification of STP temperature measurements (Fig. 1(b) of the manuscript). This microdevice featured a very stiff suspended silicon region which included a circular mesa and has a nominal thermal conductance of 240 µW.K⁻¹. Details on the fabrication and design of this device can be found in Ref. [5] This device includes a 4 kΩ Pt heater/thermometer on the suspended region. This Pt heater/thermometer was placed in a half Wheatstone bridge circuit driven by a current source (see Ref. [4] for more details). This circuit allowed passing a current through the heater for DC heating and simultaneously measuring the heater resistance which was then used for sensing the temperature of the microdevice.

## S5. Thermal Analysis

Based on the equivalent thermal circuit of Fig. 1 (e) the following first order ordinary differential equation can be written to obtain $T_{tip}$ as a function of time[6].

$$\dot{q}_{probe}(t) = \frac{T_{tip}(t)-T_R}{R_C} + \frac{T_{tip}(t)-T_S}{R_{TS}} + C_{probe}\frac{dT_{tip}(t)}{dt} \qquad \text{Eq. S1}$$

For convenience we define the total thermal resistance of the probe to the environment as:

$$\frac{1}{R_{tot}} = \frac{1}{R_C} + \frac{1}{R_{TS}} \qquad \text{Eq. S2}$$

Now consider the following scenario, the total heat deposited on the probe is the sum of constant heat flux delivered by AFM readout laser and heat flux from integrated Pt heater on the probe. By sourcing an AC current at frequency $f_1$ to the probe heater an AC heat flux at angular frequency $\omega = 4\Pi f_1$ can be generated via Joule heating. Therefore, the total probe heat flux would have a DC and AC component as shown below:

$$\dot{q}_{probe}(t) = \dot{q}_{DC} + \dot{q}_{AC}\cos(\omega t) \qquad \text{Eq. S3}$$

When combined with Eq. S1, this heat input generates a DC and AC temperature change at the probe tip with a phase lag $\varphi$ as shown below:

$$T_{tip}(t) = T_{DC} + T_{AC}\cos(\omega t + \varphi) \qquad \text{Eq. S4}$$

Now consider an experiment, where the probe is initially retracted from the surface so that there is no contact and $R_{TS} \to \infty$ (i.e. $R_{TS} \gg R_C$), and then the probe approaches the surface establishing contact. We label the DC and AC probe temperatures as $T_{DC\text{-}o}$ and $T_{AC\text{-}o}$ when the probe is out of contact (or open), and $T_{DC\text{-}c}$ and $T_{AC\text{-}c}$ when the probe is in contact with the sample. We note that, since the electrical resistance of the probe heater ($R_{heater}$) is temperature-dependent, the heat deposited on the probe by the heater ($R_{heater}I^2$) also change when its temperature is varied. Therefore, we introduce the probe heat fluxes $\dot{q}_{DC\text{-}o}$ and $\dot{q}_{AC\text{-}o}$ when the probe is out of contact (open) and $\dot{q}_{DC\text{-}c}$ and $\dot{q}_{AC\text{-}c}$ when the probe is in contact with the sample. Substituting Eq. S3 and Eq. S4 in Eq. S1, the following equation can be obtained.

For contact condition;

$$T_{DC\text{-}c} = R_{tot}\left(\dot{q}_{DC\text{-}c} + \frac{T_R}{R_C} + \frac{T_S}{R_{TS}}\right) \qquad \text{Eq. S5}$$

$$T_{AC\text{-}c} = \frac{R_{tot}\,\dot{q}_{AC\text{-}c}}{\sqrt{1+(R_{tot}\,C_{probe}\,\omega)^2}} \qquad \text{Eq. S6}$$

$$\varphi_c = -\tan^{-1}(R_{tot}\,C_{probe}\,\omega) \qquad \text{Eq. S7}$$

For out of contact condition;

$$T_{DC\text{-}o} = T_R + R_C\,\dot{q}_{DC\text{-}o} \qquad \text{Eq. S8}$$

$$\varphi_o = -\tan^{-1}(R_C\,C_{probe}\,\omega) \qquad \text{Eq. S9}$$

$$T_{AC\text{-}o} = \frac{R_C\,\dot{q}_{AC\text{-}o}}{\sqrt{1+(R_C\,C_{probe}\,\omega)^2}} \qquad \text{Eq. S10}$$

Using Eq. S5 and Eq. S8 the surface temperature can be obtained as:

$$T_S = \frac{R_{TS}}{R_C}(T_{DC\text{-}c} - T_{DC\text{-}o}) + T_{DC\text{-}c} + R_{TS}(\dot{q}_{DC\text{-}o} - \dot{q}_{DC\text{-}c}) \qquad \text{Eq. S11}$$

To obtain $R_{TS}$ we start by calculating the ratio of $T_{AC\text{-}o}/T_{AC\text{-}c}$ as shown below. In addition, for generality we include the temperature dependence of probe thermal resistance ($R_c$) and lumped thermal capacitance ($C_{probe}$).

$$\frac{T_{AC\text{-}o}}{T_{AC\text{-}c}} = \frac{R_C(T_{DC\text{-}o})\,\dot{q}_{AC\text{-}o}}{R_{tot}(T_{DC\text{-}c})\,\dot{q}_{AC\text{-}c}}\sqrt{\frac{1+R_{tot}(T_{DC\text{-}c})^2 C_{probe}(T_{DC\text{-}c})^2 \omega^2}{1+R_C(T_{DC\text{-}o})^2 C_{probe}(T_{DC\text{-}o})^2 \omega^2}} \qquad \text{Eq. S12}$$

It can be seen that at the limit of fast excitation ($\omega \gg 1/(R.C)$), the AC temperature signal becomes invariant of thermal contact resistance; therefore, this technique is only useful at low and intermediate frequencies.

Slow excitation: $\tau\omega \ll 1$, $\quad \dfrac{T_{AC\text{-}o}}{T_{AC\text{-}c}} \cdot \dfrac{\dot{q}_{AC\text{-}c}}{\dot{q}_{AC\text{-}o}} \approx \dfrac{R_c(T_{DC\text{-}o})}{R_{tot}(T_{DC\text{-}c})}$,

Fast excitation: $\tau\omega \gg 1$, $\quad \dfrac{T_{AC\text{-}o}}{T_{AC\text{-}c}} \cdot \dfrac{\dot{q}_{AC\text{-}c}}{\dot{q}_{AC\text{-}o}} \approx \dfrac{C_{probe}(T_{DC\text{-}c})}{C_{probe}(T_{DC\text{-}o})}$.

For intermediate excitation frequencies the following quadratic equation must be solved to obtain $R_{tot}$ and then $R_{TS}$.

$$R_{tot} = \sqrt{\frac{R_C(T_{DC\text{-}o})^2}{r^2\left(1+R_C(T_{DC\text{-}o})^2 C_{probe}(T_{DC\text{-}o})^2 \omega^2\right) - R_C(T_{DC\text{-}o})^2 C_{probe}(T_{DC\text{-}c})^2 \omega^2}} \qquad \text{Eq. S13}$$

Where $r$ is the ratio of AC temperature amplitudes (normalized by AC heat input) when the probe is out of contact, to contact condition, and is given by:

$$r = \frac{T_{AC\text{-}o}}{T_{AC\text{-}c}} \cdot \frac{\dot{q}_{AC\text{-}c}}{\dot{q}_{AC\text{-}o}} \qquad \text{Eq. S14}$$

After calculation of $R_{tot}$, the value of $R_{TS}$ can be readily from Eq. S2 and subsequently used for calculation of $T_S$ via Eq. S11.

## S6. Analysis of Thermal Contact Resistance

In order to compare the measured values of $R_{TS}$ with theoretical expectations, we estimated $R_{TS}$ from the following equation that was used in past work by us[7] and others[8]

$$R_{TS} = \frac{1}{\pi a} \cdot \frac{2k_s + k_{tip}\,tan\,\theta}{k_s\,k_{tip}\,tan\,\theta} \qquad \text{Eq. S15}$$

Here, $k_{tip}$ (93 Wm$^{-1}$K$^{-1}$ for chromium[9]) and $k_s$ (130 Wm$^{-1}$K$^{-1}$ for silicon[10]) are the thermal conductivities of the STP tip and sample, respectively. The contact diameter, $a$ = 5.5 nm, was obtained from Hertzian theory[1] for a tip radius of 50 nm, tip angle of $\theta$ = 45° and contact force of 50 nN. The estimated value of $R_{TS}$ is 1.68×10$^6$ K·W$^{-1}$, which is indeed in the range of experimentally observed values.